\documentclass[aps,prl,twocolumn,preprintnumbers,amsmath,amssymb]{revtex4}

\usepackage{graphicx}
\usepackage{dcolumn}
\usepackage{bm}
\usepackage{float}
\usepackage{amsfonts,amsmath,amssymb}
\usepackage{pst-grad,color}
\usepackage[bookmarks]{hyperref}

\newcommand{\up}{\uparrow}
\newcommand{\down}{\downarrow}
\def\normord#1{\ensuremath{\mathinner{
    \mathopen{\boldsymbol:}#1\mathclose{\boldsymbol:}}}}
\def\commutator#1#2{\ensuremath{\mathinner{
    \mathopen[#1,#2\mathclose]}}}
\def\anticommutator#1#2 {\ensuremath{\mathinner{
      \left\lbrace #1,#2 \right\rbrace }}}

\def\C{\mathcal{C}}

\def\pd#1#2{\frac{\partial#1}{\partial #2}} 

\begin{document}

\title{Interaction Quench in the Hubbard model}

\author{Michael Moeckel}

\author{Stefan Kehrein}

\affiliation{
Arnold-Sommerfeld-Center for Theoretical Physics and CeNS, Department Physik,\\  
Ludwig-Maximilians-Universit\"at M\"unchen, Germany.
}

\date{April 28, 2008}

\begin{abstract} 
Motivated by recent experiments in ultracold atomic gases that explore the nonequilibrium dynamics
of interacting quantum many-body systems, we investigate the opposite limit of Landau's Fermi 
liquid paradigm: We study a Hubbard model with a sudden interaction quench, that is the 
interaction is switched on at time $t=0$. Using the flow equation method, we are able to study the
real time dynamics for weak interaction~$U$ in a systematic expansion and find three clearly separated
time regimes: i)~An initial buildup of correlations where the quasiparticles are formed. ii)~An intermediate
quasi-steady regime resembling a zero temperature Fermi liquid with a nonequilibrium quasiparticle
distribution function.
iii)~The long time limit described by a quantum Boltzmann equation 
leading to thermalization of the momentum distribution function with a temperature~$T\propto U$.
\end{abstract}

% \pacs{Valid PACS appear here}%
\maketitle

The investigation of interacting quantum many-particle systems in nonequilibrium has recently attracted a lot of attention. 
A simple way to excite a system from its ground state is an interaction quench, 
a sudden switch of parameters in the Hamiltonian. The time evolution of the initial state
is then generated by the quenched Hamiltonian, for which the initial state is generically not an eigenstate.
Recent experiments have implemented quenches of ultracold atoms loaded on optical lattices and observed remarkable subsequent dynamics described as iterated 'collapse and revival' of the 
initial superfluid phase \cite{Greiner2002_2, Kinoshita2006}. 
Yet their theoretical description remains a challenge since many well-established equilibrium theoretical methods fail in nonequilibrium. 
From a theoretical point of view, the long-time limit poses particularly intriguing questions: Will an interacting closed quantum system prepared in some generic initial state equilibrate, that is behave like the equilibrium system with some nonzero temperature after waiting sufficiently long?  
In nonlinear classical systems similar questions have been addressed in a multitude of publications since the seminal work by Fermi, Pasta and Ulam \cite{Ford1992}. 
Non-equilibration has been linked to integrability since an integrable system is constrained by an infinite number
of conservation laws. 

However, much less is known about quantum systems. 
Since a pure state remains a pure state under unitary time evolution, the concept of thermalization
is only meaningful for suitable observables.
First theoretical results
have shown that observables may approach limiting values or exhibit persistent oscillations which, 
even when time-averaged, do not match with equilibrium properties \cite{Altshuler2006,Leggett2005}.
A proposition by Rigol et al. \cite{Rigol2007} gave a statistical description 
for the stationary state of an integrable system in terms of a generalized Gibbs ensemble. 
Conditions for the applicability or non-applicability of this scenario have been clarified in \cite{Barthel2008} 
and specific results have been obtained 
for the Luttinger model \cite{Cazalilla2006}, hard core bosons in one dimension \cite{Rigol2006A,Pustilnik2007}
and the infinite dimensional Falicov-Kimball model \cite{Eckstein2007}. 
While the concept of a generalized statistical ensemble proved helpful even for a less restrictive set of 
constraints  \cite{Manmana2007},  the role of integrability has been questioned by further numerical works: 
Breaking the integrability of spinless fermions on a 1d lattice has not altered relaxation to a non-thermal 
state \cite{Manmana2007}. Similarly, for the non-integrable 1d Bose-Hubbard model signatures of thermalization 
could only be found for a limited regime of quenches, while others seemed to drive the system to non-thermal 
stationary states \cite{Kollath2007}. Exact results have been obtained for the opposite case of quenches
from the Mott phase to the noninteracting Hamiltonian and show relaxation of local observables to
a nonequilibrium steady state \cite{Eisert2008}.

Motivated by these questions, 
we study an interaction quench in a Fermi liquid in $d>1$ spatial dimensions, that is we suddenly switch on the interaction at time $t=0$. This is the extreme opposite limit of Landau's adiabatic switching on
procedure, where one finds the celebrated one to one mapping between physical electrons and quasiparticles.
In the sudden quench scenario, the system is prepared as the zero temperature ground state
of the noninteracting Fermi gas at times $t<0$, and then, for $t\geq 0$, subject to the time evolution with respect to 
the interacting Hamiltonian. 
We find three regimes of the time evolution, that are well separated for
weak interaction: An initial quasiparticle formation regime, followed by a quasi-steady intermediate regime resembling
a zero temperature Fermi liquid, and a long-time thermalization regime where the momentum distribution function equilibrates. 
Concretely, we investigate the fermionic Hubbard model at half filling 
described by the following Hamiltonian (Fermi energy $\epsilon_{F}\equiv 0$)
\begin{equation}
H(t) =
\sum_{k\sigma=\up,\down}
\epsilon_k :c^{\dagger}_{k\sigma} c^{\phantom\dagger}_{k\sigma}:
+\Theta(t)\,U\,
 \sum_{i} (n_{i\up}-\frac{1}{2})
(n_{i\down}-\frac{1}{2})
\label{defHt}
\end{equation}
and work out the 
time-dependent momentum distribution functions $N_{k}(t)$. 
Notice that this system is clearly non-integrable for $d>1$ and one therefore expects generic behavior.
Most of our results are obtained in the limit of high dimensions \cite{Vollhardt1992}, but the calculation
also applies to finite dimensions with the same conclusions up to quantitative details. 

 \begin{figure}
 \begin{center}
  \includegraphics[height=25mm]{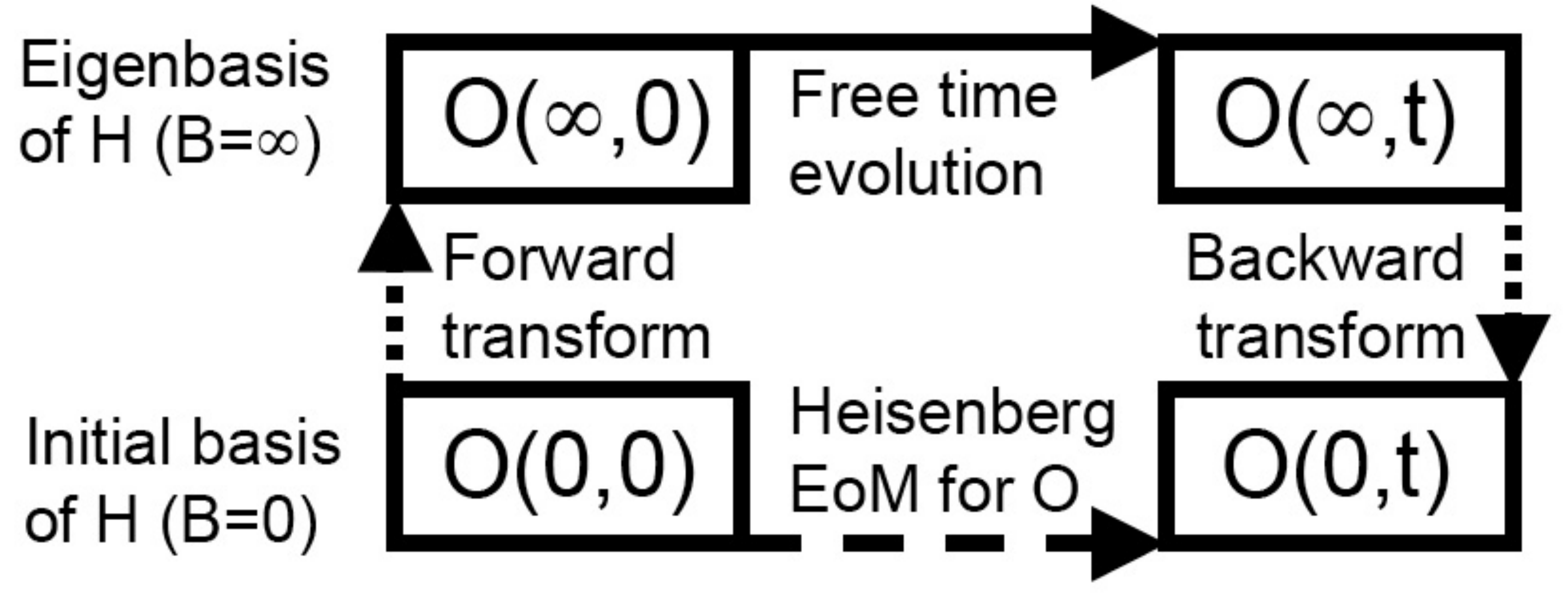} 
    \caption{The Heisenberg equation of motion for an observable $O$ is solved by transforming to 
    the $B=\infty$ eigenbasis of the interacting 
    Hamiltonian $H$ (forward transformation), where the time evolution can be computed easily.
   Time evolution introduces phase shifts, and therefore the form of the observable in the initial
   basis $B=0$ (after a backward transformation) changes as a function of time.}
  \label{Fig_Rechenschema}
  \end{center}
 \end{figure}
We study the above real time evolution problem by using the approach introduced in~\cite{Hackl2007}. 
One solves the Heisenberg equations of motion for the operators that one is interested in 
by performing a unitary transformation to an (approximate) eigenbasis of the interacting Hamiltonian. 
There one can easily work out the time evolution and then transform back to the original basis
where the initial state is specified. In this manner one induces a solution of the Heisenberg equations
of motion for an operator in the original basis but without secular terms, which are usually a major problem
in other approximation schemes \cite{Berges2005}. Fig. \ref{Fig_Rechenschema} gives a sketch of our approach.
Notice that the same general idea was recently also used by
Cazalilla to study the behavior of the exactly solvable one-dimensional Luttinger model subject to a quench \cite{Cazalilla2006}.

Since our model is non-integrable, 
we implement the above diagonalizing transformation by the flow equation method \cite{Wegner1994,Kehrein_book},
which permits a systematic controlled expansion for many equilibrium and nonequilibrium 
quantum many-body problems  \cite{Kehrein_book}.
One uses a continuous sequence of infinitesimal unitary transformations 
parametrized by a parameter $B$ with dimension $(\text{energy})^{-2}$ that connects the eigenbasis of the free 
Hamiltonian ($B=0$) with the energy diagonal basis of the interacting Hamiltonian 
($B=\infty$). Each infinitesimal step of the unitary transformation is defined by 
the canonical generator $\eta(B) = \commutator{H_0(B)}{H_{\text{int}}(B)}$, 
where $H_{0}(B)$ is the diagonal and $H_{\text{int}}(B)$ the interacting part of the Hamiltonian.
This generator $\eta(B)$ has the 
required property of making $H(B)$ increasingly energy diagonal for $B\rightarrow \infty$ \cite{Wegner1994}.
All operators $\mathcal{O}(B)$ (including the Hamiltonian itself) flow according to the 
differential equation $\partial{\mathcal{O}(B)} / \partial B = \commutator{\eta(B)}{\mathcal{O}(B)}$. 
Higher order terms generated by the commutator are truncated after normal-ordering (denoted by $:\; :$) and
the flow equations  decompose into a set of ordinary differential equations resembling scaling equations in a renormalization approach.  
However, contrary to conventional renormalization schemes which reduce the size of the effective Hilbert space, 
the flow equation approach retains the full Hilbert space, which makes it particularly appropriate for
nonequilibrium problems (for more details see \cite{Kehrein_book}).

\emph{Flow equations for the Hubbard model.}
First we work out the
diagonalizing flow equation transformation for the Hubbard Hamiltonian. 
The expansion parameter is the (small) interaction $U$
and normal-ordering is 
with respect to the zero temperature Fermi-Dirac distribution:
\begin{eqnarray}
H(B) &=& 
\sum_{k\sigma=\up,\down}
\epsilon_k \normord{c^{\dagger}_{k\sigma}c^{\phantom\dagger}_{k\sigma}} 
\\ &+& \nonumber
 \sum_{p'pq'q} 
 U_{p'pq'q}(B) \normord{c^{\dagger}_{p'\up}c^{\phantom\dagger}_{p\up} c^{\dagger}_{q' \down}c^{\phantom\dagger}_{q\down}} %\\
\end{eqnarray}
with $U_{p'pq'q}(B=0)=U$.
The flow of the one-particle energies and the generation of higher normal-ordered terms in the Hamiltonian 
can be neglected since we are interested in results in second order in~$U$.
The flow of the interaction is to leading order given by
$U_{p'pq'q}(B) = U  \exp({-B\Delta_{p'pq'q}^2})$  
with an energy difference $\Delta_{p'pq'q} \stackrel{\rm def}{=} \epsilon_{p'} - \epsilon_p + \epsilon_{q'}-\epsilon_{q}$.

Next we work out the flow equation transformation for the number operator 
$\mathcal{N}_{k \up} (B) = \C^{\dagger}_{k \up}(B) \, \C^{\phantom\dagger}_{k \up}(B) $, which 
can be obtained from the transformation of a single creation operator $\C_{k\up}^{\dagger}(B)$. 
Under the sequence of unitary transformations the operator changes its form to describe
dressing by electron-hole pairs. A truncated ansatz reads:
\begin{equation}
\label{Ansatz-C}
\C^{\dagger}_{k \up}(B) \hspace{-2pt}=\hspace{-2pt}   h_{k }(B) c^{\dagger}_{k \up} 
+ \hspace{-3pt}
\sum_{p'q'p} M^{k}_{p'q'p}(B) \delta^{k+p}_{p'+q'} 
\normord{c^{\dagger}_{p'\up} c^{\dagger}_{q' \down} c^{\phantom\dagger}_{p \down}} 
\end{equation}
We introduce the zero temperature momentum distribution function of a free Fermi gas $n_k$, 
define $n_k^{-} \stackrel{\rm def}{=} 1-n_k$ and a phase space factor  
$Q_{p'pq'}[n] \stackrel{\rm def}{=} n^{-}_{p'}n^{-}_{q'}n^{}_p+n^{}_{p'}n^{}_{q'}n^{-}_p$. 
The flow equations for the creation operator are:  
\begin{multline}
\pd{h_{k}(B)}{B} = U \sum_{p'q'p} M^{k}_{p'q'p}(B) \ \Delta_{kp'pq'}  \,
e^{-B\Delta_{kp'pq'}^2} \  Q_{p'pq'}[n] 
\\ 
\pd{M_{p'q'p}^{k}(B)}{B} = h_{k}(B)\, U \, \Delta_{p'pq'k} e^{-B\Delta_{p'pq'k}^2} \qquad ~
\label{FGL-M}
\end{multline}
Here and in the ansatz (\ref{Ansatz-C}) we have only taken into account the terms that are required to describe the momentum distribution function up to second order in~$U$.
The initial conditions for the above transformation of $\C^{\dagger}_{k\up}$ are $h_k(0)=1$ and 
\mbox{$M^{k}_{p'q'p}(0)=0$} (i.e., $\C^{\dagger}_{k\up}(B=0)=c^{\dagger}_{k\up}$),
and we denote the asymptotic values from the solution of (\ref{FGL-M})
by $h_{k}(B=\infty,t=0)$ and $M^{k}_{p'q'p}(B=\infty,t=0)$. Time evolution according to
Fig.~\ref{Fig_Rechenschema} yields $h_{k}(B=\infty,t)=h_{k}(B=\infty,t=0)\,e^{-i\epsilon_{k}t}$
and  $M^{k}_{p'q'p}(B=\infty,t)=M^{k}_{p'q'p}(B=\infty,t=0)\,e^{-i(\epsilon_{p'}+\epsilon_{q'}-\epsilon_{p})t}$,
which are then input as the initial conditions of the system of equations (\ref{FGL-M}) at $B=\infty$.
Integrating back to $B=0$ gives the time evolved creation operator in the original basis, and it is 
straightforward to evaluate
the time dependent momentum distribution function with respect to the initial Fermi gas state \cite{Details}.

\begin{figure}
  \begin{center}
      \includegraphics[width=85mm]{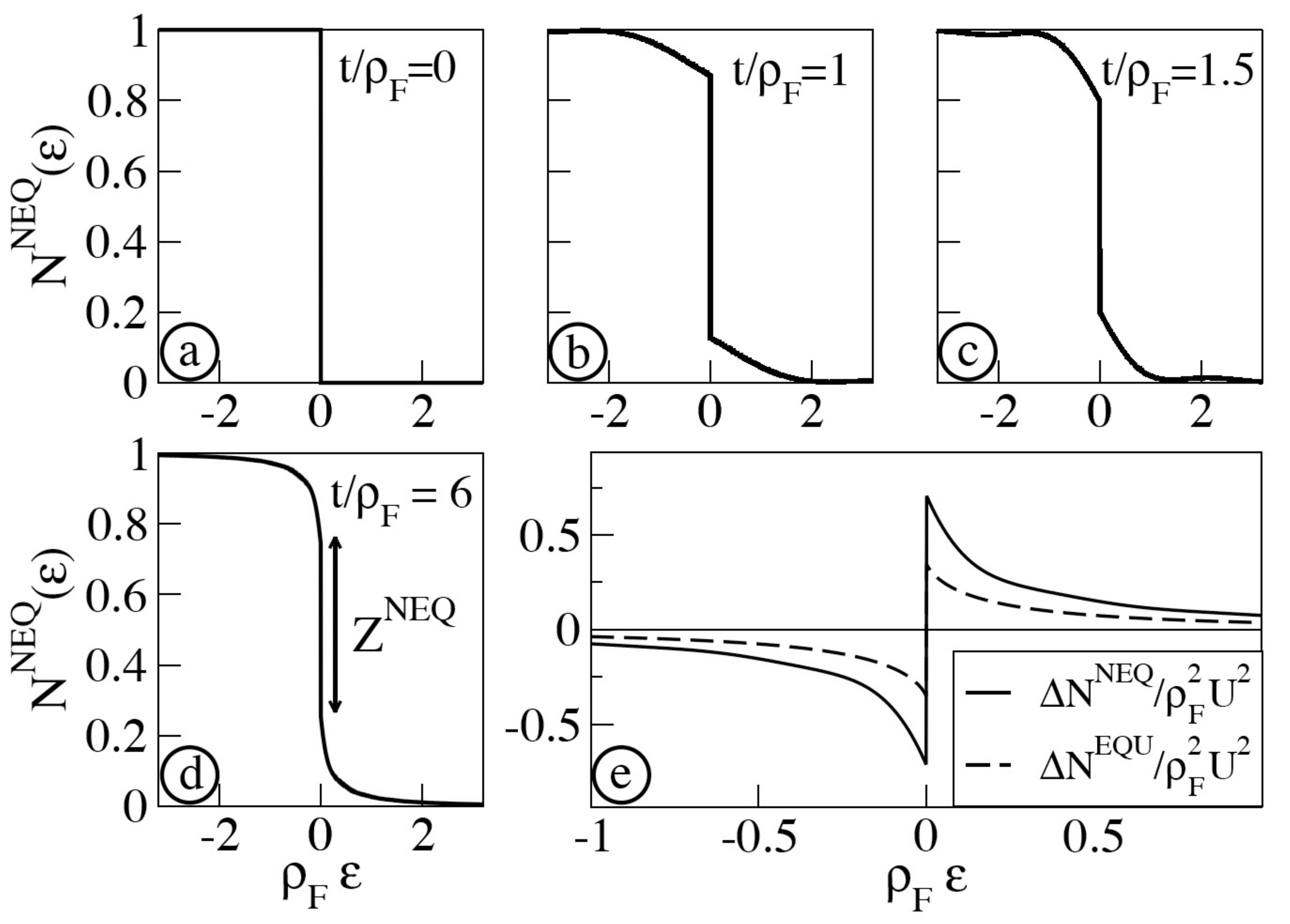}   \hspace{-0.5cm}
    \caption{(a)-(d): Time evolution of 
    $N^{\rm NEQ}(\epsilon)$ plotted around the Fermi energy for $\rho_F U=0.6$. A fast reduction of the discontinuity and 1/t-oscillations 
    can be observed. The arrow in (d) indicates the size of the quasiparticle residue in the 
    quasi-steady regime. In (e) the universal curves for
    $\Delta N_{k} = N_k -n_k$ are given for both equilibrium and for the nonequilibrium 
    quasi-steady state in the weak-coupling limit.}
    \label{Fig_NGG_FDF}
  \end{center}
\end{figure}

\emph{Nonequilibrium momentum distribution function.}
One finds the following time-dependent additional term to the distribution $n_{k}$ of the
free Fermi gas in~$O(U^{2})$:
\begin{eqnarray}
\Delta N^{\rm NEQ}_{k}(t)&=&
N^{\rm NEQ}_{k}(t) - n_{k} \label{NGG-FDF} \\
&=& - 4U^2 \int_{-\infty}^{\infty} dE {\frac{\sin^2\left( \frac{(\epsilon_{k}-E)t}{2}\right)}{(\epsilon_{k}-E)^2}} \ J_k(E;n)
\nonumber 
\end{eqnarray}
The phase space factor $J_k(E;n) $ resembles the quasiparticle collision integral of a quantum  Boltzmann equation: 
\begin{equation*}
J_k(E;n) = \sum_{p'q'p} \delta^{p'+q'}_{p+k} \delta^{\epsilon_{p'}+\epsilon_{q'}}%-
_{\epsilon_p +E} %) \\
\left[ n_{k}  n_p n^-_{p'} n^-_{q'}  - n_k^- n_p^- n_{p'} n_{q'} \right] 
\label{PSF_J-Def}
\end{equation*}
For computational convenience we use the 
limit of infinite dimensions, specifically a Gaussian density of
states $\rho(\epsilon)=\exp\left(-(\epsilon/t^{*})^{2}/2\right)/\sqrt{2\pi}t^{*}$ 
\cite{Vollhardt1992}. In the sequel $\rho_{F}=\rho(\epsilon=0)$ denotes the density
of states at the Fermi level. 
Results from a numerical evaluation of the above scheme for three time steps are presented in Fig.~\ref{Fig_NGG_FDF}.

\emph{Equilibrium momentum distribution function.}
Eqs.~(\ref{FGL-M}) can also be used to evaluate the equilibrium 
distribution function, which will later be important for comparison. In fact, the asymptotic
value $h_{k_{F}}(B=\infty)$ at the Fermi energy is directly related to the quasiparticle
residue ($Z$-factor), $Z^{\rm EQU}=[h_{k_{F}}(B=\infty)]^{2}$ \cite{Kehrein_book}.
It is easy to solve (\ref{FGL-M}) analytically at the Fermi energy for zero
temperature in~$O(U^{2})$ and one finds for momenta~$k$ infinitesimally above or below the
Fermi surface
\begin{equation}
\Delta N^{\rm EQU}_{k}
= -U^2 \int_{-\infty}^{\infty} dE\, \frac{J_k(E;n)}{(\epsilon_{k}-E)^2}
\end{equation}
consistent with a conventional perturbative evaluation.

\emph{Short-time correlation buildup.}
The numerical evaluation of the momentum distribution function depicted in Fig.~\ref{Fig_NGG_FDF}
shows the initial buildup of a correlated state from the Fermi gas. For times $0< t\lesssim \rho_F^{-1} U^{-2}$
one observes a fast reduction of the Fermi surface discontinuity and $1/t$ oscillations in the momentum distribution 
function. This short time regime can be understood as the formation of quasiparticles from the free electrons of
the initial noninteracting Fermi gas.

\emph{Intermediate quasi-steady regime.}
For times $t$ of order $\rho_F^{-1} U^{-2}$ the sinusoidal time dependence in (\ref{NGG-FDF}) generates an increasing 
localization in energy space, which eventually becomes a $\delta$-function (Fermi's golden rule). 
There are no further changes in the momentum distribution function for times $t\gtrsim \rho_F^{-1} U^{-2}$
in the present order of the calculation. For momenta~$k$ infinitesimally above
or below the Fermi surface one then finds from (\ref{NGG-FDF}):
\begin{eqnarray}
\Delta N^{\text{NEQ}}_{k}(t\rightarrow \infty)&=& 
-4U^2 \int_{-\infty}^{\infty} dE\, \frac{1}{2}\,\frac{J_k(E;n)}{(\epsilon_{k}-E)^2}
\nonumber \\ 
&=& 2 \, \Delta N_{k}^{\text{EQU}}
\label{eq_factortwo}
\end{eqnarray}
since $\sin^{2}$ in (\ref{NGG-FDF}) yields a factor~$1/2$ in the long time limit.
In the quasi-steady state the momentum distribution function is therefore that of a
zero temperature Fermi liquid. However, from (\ref{eq_factortwo}) one deduces that its $Z$-factor is smaller than in equilibrium,
$1-Z^{\rm NEQ}=2(1-Z^{\rm EQU})$. This factor~2 implies a quasiparticle distribution function in the
vicinity of the Fermi surface in the quasi-steady
state equal to the equilibrium distribution function of the physical electrons,
$N^{\rm QP:NEQ}_{k}=N^{\rm EQU}_{k}$, as opposed to its equilibrium distribution, 
$N^{\rm QP:EQU}_{k}=\Theta(k_{F}-k)$.

Remarkably, Cazalilla's findings \cite{Cazalilla2006} for the interaction quench in the Luttinger model 
mirror these features: the critical exponent describing the asymptotic behavior of the electronic
Green's function differs from the equilibrium result.
As Cazalilla points out this corresponds to a non-equilibrium distribution for the bosonic modes after bosonization.
A main difference between the Luttinger liquid and the Fermi liquid cases follows from the
integrability of the Luttinger liquid with an infinite number of conservation laws,
which make this regime stable for $t\rightarrow\infty$. For the Fermi liquid,
on the other hand, on-shell interactions lead to thermalization as we will see next.

\emph{Thermalization.}
The previous flow equation calculation of the real time dynamics contains all contributions to the 
time evolution for times smaller than $\rho_F^{-3} U^{-4}$.
For the long time dynamics one generally expects a quantum Boltzmann equation (QBE) to be a 
valid description \cite{Rammer1986}
\begin{equation}
\pd{N^{\rm QP}_k(t)}{t} = -  \rho_{F}\,U^2\,J_{k}(E=\epsilon_{k},N^{\rm QP}(t)) \ .
\label{QBE}
\end{equation}
Here the quasiparticle momentum distribution function $N^{\rm QP:NEQ}_{k}$ derived above serves as the 
initial condition. Because $N^{\rm QP:NEQ}_{k}$ allows nonzero phase space for scattering processes
in the vicinity of the Fermi surface (originating, ultimately, from the factor~2 in (\ref{eq_factortwo})),
the initial quasiparticle distribution function starts to evolve on the time scale $t\propto \rho_F^{-3} U^{-4}$.
This implies that the quasi-steady electron distribution function depicted in Fig.~\ref{Fig_NGG_FDF}d
starts to decay on this time scale and one approaches
a Fermi-Dirac distribution (being the only stable fixed point of (\ref{QBE}))
with a nonzero temperature~$T$.

The above scenario fits well into the picture of nonequilibrium field theories describing, e.g., 
the early universe \cite{Berges2005}. 
The excitation energy of the initial quantum state (the Fermi gas) with respect to the equilibrium
ground state of (\ref{defHt}) is $E_{\rm ex}=\alpha\rho_{F}U^{2}$ in the weak interaction limit with some
lattice-dependent constant~$\alpha>0$. 
The short-time correlation buildup corresponds to prethermalization, where kinetic and interaction
energy in (\ref{defHt}) flow from 0 to $E^{\rm NEQ}_{\rm int}=-2\alpha\rho_{F}U^{2}$
and  $E^{\rm NEQ}_{\rm kin}=2\alpha\rho_{F}U^{2}$. This follows immediately from
the Feynman-Hellman theorem and the fact that the total energy remains zero 
for all times. $E^{\rm NEQ}_{\rm int}$ equals the equilibrium interaction energy, while 
$E^{\rm NEQ}_{\rm kin}=E^{\rm EQU}_{\rm kin}+E_{\rm ex}$.
Kinetic and interaction energy then remain constant throughout the quasi-steady regime and the 
long-time limit, and therefore the system has {\em prethermalized\/} for these average quantities. 
In the thermalization regime the system redistributes its additional excitation energy $E_{\rm ex}$
in the kinetic energy over the different momenta and reaches a Fermi-Dirac distribution
with temperature $T\propto U$.

\emph{Higher order flow equations.}
Clearly, it would be desirable to derive  (\ref{QBE}) within the framework of the real time flow equation
calculation. However, a calculation to order~$U^{4}$ is
beyond the scope of the present work. Still, one can identify a particular contribution in fourth order
leading to a finite lifetime of order $\rho_{F}^{-3}U^{-4}$ for an electron at the Fermi surface,
which is consistent with the dynamics implied by the
QBE. The short time evolution of the system for times smaller than $\rho_{F}^{-3}U^{-4}$ obtained
from the full solution of the Heisenberg equations of motion therefore matches the long time
dynamics described by the QBE, and we have a consistent picture on all time scales.
Another effect of the fourth order contributions is that the sharp Fermi edge of the quasi-steady
state gets smeared out on an energy scale $\rho_{F}^{3}U^{4}$, which, however, does not essentially modify
our previous conclusions. Therefore, strictly speaking the discontinuity of the momentum
distribution function disappears immediately for $t>0$, but this effect only becomes noticeable
for times of order~$\rho_{F}^{-3}U^{-4}$.

\emph{Conclusions.}
We have discussed the real time evolution of the Hubbard model with a sudden interaction quench for
a weak interaction ~$U$. Ultimately, the system completely thermalizes its excitation energy~$E_{\rm ex}$ and reaches 
a temperature $T\propto U$. This thermalization
regime only sets in on the time scale~$\rho_{F}^{-3}U^{-4}$. This follows from the observation
that the short time behavior up to times of order~$\rho_{F}^{-1}U^{-2}$ amounts to quasiparticle formation
with a momentum distribution function with a discontinuity at the Fermi energy. Therefore, a quasi-steady prethermalized
state emerges for times $\rho_{F}^{-1}U^{-2}\lesssim t\lesssim \rho_{F}^{-3}U^{-4}$.
Its momentum distribution function looks like a zero temperature Fermi liquid, but with
the wrong quasiparticle residue with respect to the interacting ground state. It is 
this nonequilibrium quasiparticle residue that allows
for phase space for scattering processes in a quantum Boltzmann equation description for
times $t\gtrsim \rho_{F}^{-3}U^{-4}$, which then leads to thermalization of the momentum distribution function.

We acknowledge valuable discussions with K.~Morawetz and F.~Marquardt. This work
was supported through SFB~631
of the Deutsche Forschungsgemeinschaft, the
Center for Nanoscience (CeNS) Munich and the German
Excellence Initiative via the Nanosystems Initiative Munich (NIM).


\vspace*{-0.5cm}
\begin{thebibliography}{18}
\vspace*{-0.5cm}
\expandafter\ifx\csname natexlab\endcsname\relax\def\natexlab#1{#1}\fi
\expandafter\ifx\csname bibnamefont\endcsname\relax
  \def\bibnamefont#1{#1}\fi
\expandafter\ifx\csname bibfnamefont\endcsname\relax
  \def\bibfnamefont#1{#1}\fi
\expandafter\ifx\csname citenamefont\endcsname\relax
  \def\citenamefont#1{#1}\fi
\expandafter\ifx\csname url\endcsname\relax
  \def\url#1{\texttt{#1}}\fi
\expandafter\ifx\csname urlprefix\endcsname\relax\def\urlprefix{URL }\fi
\providecommand{\bibinfo}[2]{#2}
\providecommand{\eprint}[2][]{\url{#2}}

\bibitem[{\citenamefont{{K}inoshita et~al.}(2006)\citenamefont{{K}inoshita,
  {W}enger, and {W}eiss}}]{Kinoshita2006}
\bibinfo{author}{\bibfnamefont{T.}~\bibnamefont{{K}inoshita}},
  \bibinfo{author}{\bibfnamefont{T.}~\bibnamefont{{W}enger}}, \bibnamefont{and}
  \bibinfo{author}{\bibfnamefont{D.}~\bibnamefont{{W}eiss}},
  \bibinfo{journal}{Nature  (London)} \textbf{\bibinfo{volume}{440}},
  \bibinfo{pages}{900} (\bibinfo{year}{2006}).

\bibitem[{\citenamefont{M.~Greiner and Bloch}(2002)}]{Greiner2002_2}
\bibinfo{author}{\bibnamefont{M.~Greiner}} {\it et al.},
 % \bibfnamefont{O.~Mandel} \bibnamefont{ T.~H\"ansch, and I.~Bloch}},
 % \bibinfo{author}{\bibfnamefont{I.}~\bibnamefont{Bloch}},
  \bibinfo{journal}{Nature (London)} \textbf{\bibinfo{volume}{419}}, \bibinfo{pages}{51}
  (\bibinfo{year}{2002}).

\bibitem[{\citenamefont{Ford}(1992)}]{Ford1992}
\bibinfo{author}{\bibfnamefont{J.}~\bibnamefont{Ford}}, \bibinfo{journal}{Phys.
  Rep.} \textbf{\bibinfo{volume}{213}}, \bibinfo{pages}{271}
  (\bibinfo{year}{1992}).

\bibitem{Altshuler2006}
E. A. Yuzbashyan, O. Tsyplyatyev, and B. L. Altshuler,
Phys. Rev. Lett. {\bf 96}, 097005 (2006).

\bibitem{Leggett2005}
G. L. Warner and A. J. Leggett, Phys. Rev. B {\bf 71}, 134514 (2005).

\bibitem[{\citenamefont{M.~Rigol and Olshanii}(2007)}]{Rigol2007}
\bibinfo{author}{ \bibnamefont{M.~Rigol {\it et al.}}},
  \bibinfo{journal}{Phys. Rev. Lett.} \textbf{\bibinfo{volume}{98}}, \bibinfo{pages}{050405}
  (\bibinfo{year}{2007}).

\bibitem{Barthel2008}
T. Barthel and U. Schollw{\"o}ck, 
Phys. Rev. Lett. {\bf 100}, 100601 (2008).

\bibitem[{\citenamefont{{C}azalilla}(2006)}]{Cazalilla2006}
\bibinfo{author}{\bibfnamefont{M.~A.} \bibnamefont{{C}azalilla}},
  \bibinfo{journal}{Phys. Rev. Lett.} \textbf{\bibinfo{volume}{97}}, \bibinfo{pages}{156403}
  (\bibinfo{year}{2006}).

\bibitem[{\citenamefont{M.~{R}igol}(2006)}]{Rigol2006A}
\bibinfo{author}{\bibnamefont{M.~{R}igol, A.~Muramatsu, and M.~Olshanii}}, \bibinfo{journal}{Phys. Rev. A}
  \textbf{\bibinfo{volume}{74}}, \bibinfo{pages}{053616}
  (\bibinfo{year}{2006}).

\bibitem{Pustilnik2007}
D. M. Gangardt and M. Pustilnik, cond-mat/0709.2374

\bibitem{Eckstein2007}
M. Eckstein and M. Kollar, Phys. Rev. Lett. {\bf 100}, 120404 (2008).

\bibitem[{\citenamefont{S.~{R}.~{M}anmana and {M}uramatsu}(2007)}]{Manmana2007}
\bibinfo{author}{ \bibnamefont{S.~{R}.~{M}anmana {\it et al.}}},
  \bibinfo{journal}{Phys. Rev. Lett.} \textbf{\bibinfo{volume}{98}}, \bibinfo{pages}{210405}
  (\bibinfo{year}{2007}).

\bibitem[{\citenamefont{C.~{K}ollath and {A}ltmann}(2007)}]{Kollath2007}
\bibinfo{author}{ \bibnamefont{C.~{K}ollath, A.~L\"auchli, and E.~Altman},}
   \bibinfo{journal}{Phys. Rev. Lett.} \textbf{\bibinfo{volume}{98}}, \bibinfo{pages}{180601}
  (\bibinfo{year}{2007}).

\bibitem{Eisert2008}
M. Cramer {\it et al.}, Phys. Rev. Lett. {\bf 100}, 030602 (2008).

\bibitem[{\citenamefont{Vollhardt}(1993)}]{Vollhardt1992}
\bibinfo{author}{\bibfnamefont{D.}~\bibnamefont{Vollhardt}},
  \bibinfo{journal}{in {\it Correlated {E}lectron {S}ystems}, edited by V.J. Emery
  (World Scientific, Singapore, 1993), p. 57}

\bibitem[{\citenamefont{A.~Hackl}(2007)}]{Hackl2007}
\bibinfo{author}{\bibnamefont{A.~Hackl and S. Kehrein}},
  \bibinfo{journal}{cond-mat/07092100}  

\bibitem[{\citenamefont{Berges}(2005)}]{Berges2005}
\bibinfo{author}{\bibfnamefont{J.}~\bibnamefont{Berges}}, \bibinfo{journal}{AIP
  Conf. Proc.} 
  \textbf{\bibinfo{volume}{739}}, \bibinfo{pages}{3}
  (\bibinfo{year}{2004}).

\bibitem[{\citenamefont{Wegner}(1994)}]{Wegner1994}
\bibinfo{author}{\bibfnamefont{F.}~\bibnamefont{Wegner}},
  \bibinfo{journal}{Ann. Phys. (Leipzig)} \textbf{\bibinfo{volume}{506}},
  \bibinfo{pages}{77} (\bibinfo{year}{1994}).

\bibitem[{\citenamefont{Kehrein}(2006)}]{Kehrein_book}
\bibinfo{author}{\bibfnamefont{S.}~\bibnamefont{Kehrein}},
  \emph{\bibinfo{title}{The {F}low {E}quation {A}pproach to {M}any-{P}article
  {S}ystems}} (\bibinfo{publisher}{Springer, Berlin}, \bibinfo{year}{2006}).

\bibitem{Details}
Details of this calculation will be presented elsewhere. Notice that our calculation cannot
be used for the $d=1$ Hubbard model since it is nonperturbative in~$U$.

\bibitem[{\citenamefont{J.~Rammer}(1986)}]{Rammer1986}
\bibinfo{author}{\bibnamefont{J.~Rammer and H.~Smith}},
  \bibinfo{journal}{Rev. Mod. Phys.} \textbf{\bibinfo{volume}{58 (2)}},
  \bibinfo{pages}{323} (\bibinfo{year}{1986}).


\end{thebibliography}
\end{document}